\documentclass[a4paper,UKenglish,cleveref, autoref, thm-restate]{lipics-v2021}

\usepackage{anyfontsize}

\usepackage{mathptmx}
\usepackage{import}
\usepackage{hyperref}
\usepackage{tabularx}
\usepackage{times}

\usepackage{tabularray}
\usepackage{siunitx}
\hideLIPIcs
\nolinenumbers
        \UseTblrLibrary{siunitx}
        \DefTblrTemplate{caption-tag}{default}{\kern0.05em{\color[rgb]{0.99,0.78,0.07}\rule{0.73em}{0.73em}}%
        \hspace*{0.67em}\bfseries Table\hspace{0.25em}\thetable}
\DefTblrTemplate{caption-sep}{default}{\bfseries:\enskip}
\DefTblrTemplate{caption-text}{default}{\InsertTblrText{caption}}
\usepackage{changepage}
\usepackage{algorithm}
\usepackage[noend]{algorithmic}
\newcommand\algorithmicprocedure{\textbf{procedure}}
\newcommand{\algorithmicendprocedure}{\algorithmicend\ \algorithmicprocedure}
\makeatletter
\newcommand\PROCEDURE[3][default]{%
  \ALC@it
  \algorithmicprocedure\ \textsc{#2}(#3)%
  \ALC@com{#1}%
  \begin{ALC@prc}%
}
\newcommand\ENDPROCEDURE{%
  \end{ALC@prc}%
  \ifthenelse{\boolean{ALC@noend}}{}{%
    \ALC@it\algorithmicendprocedure
  }%
}
\newenvironment{ALC@prc}{\begin{ALC@g}}{\end{ALC@g}}
\makeatother

\newcommand{\HRdataset}[1]{#1}
\newcommand{\HRinstance}[1]{\texttt{#1}}

\usepackage{rotating}
\usepackage{multirow}
\usepackage{numprint}

\usepackage{amsmath,mathtools}
\usepackage{showexpl}
\usepackage{wrapfig}

\newif\ifCompileTables
\CompileTablestrue

\newif\ifFullPaper
\FullPaperfalse

\DeclareRobustCommand{\frcshape}{\fontfamily{frc}\selectfont}
\DeclareTextFontCommand{\textfrc}{\frcshape}

\newtheorem{@reduction}{Reduction}[section]

\usepackage[usenames,dvipsnames]{xcolor}
\usepackage{hyperref}
\usepackage{pgf}
\usepackage{pgfplots}
\usepackage{iftex}
\usepackage{ifxetex,ifluatex}

\newif\ifunicode
\ifxetex\unicodetrue\fi
\ifluatex\unicodetrue\fi
\ifluatex

\else 

\fi
\ifPDFTeX
\usepackage[utf8]{inputenc}
\else

\ifunicode

  \usepackage{newunicodechar}
  \newcommand{\DeclareUnicodeCharacter}[2]{%
    \begingroup\lccode`|=\string"#1\relax
    \lowercase{\endgroup\newunicodechar{|}}{#2}%
  }
\else
  \usepackage[utf8]{inputenc}
\fi
\fi
\DeclareUnicodeCharacter{2212}{−}
\usepackage{tikz}
\usepgfplotslibrary{groupplots,dateplot}

\usetikzlibrary{patterns,shapes.arrows}
\pgfplotsset{compat=newest}

\tikzset{semithick/.style={line width =1.5pt}}
\tikzset{font={\small}}

\def\stripzero#1{\expandafter\stripzerohelp#1}
\def\stripzerohelp#1{\ifx 0#1\expandafter\stripzerohelp\else#1\fi}

\newcommand{\justprint}[1]{#1}

\newcommand{\cpp}{\textsf{C}\texttt{++}\xspace}

\newcommand{\Repeat}   {{\bf repeat\ }}
\newcommand{\Until}    {{\bf until\ }}

\newcommand{\Return}   {{\bf return\ }}

\newcommand{\ie}{i.\,e.,\xspace}

\newcommand{\etal}{et~al.\xspace}

\def\comment#1{}

\def\withcomments{
	\newcounter{mycommentcounter}
	\def\comment##1{\refstepcounter{mycommentcounter}%
		\ifhmode%
		\unskip%
		{\dimen1=\baselineskip \divide\dimen1 by 2 %
			\raise\dimen1\llap{\tiny\bfseries \textcolor{red}{-\themycommentcounter-}}}\fi%
		\marginpar[{\renewcommand{\baselinestretch}{0.8}%
			\hspace*{3em}\begin{minipage}{5em}\footnotesize [\themycommentcounter]: \raggedright ##1\end{minipage}}]{\renewcommand{\baselinestretch}{0.8}%
			\begin{minipage}{5em}\footnotesize [\themycommentcounter]: \raggedright ##1\end{minipage}}}
}

\acknowledgements{We acknowledge support by DFG grant SCHU 2567/3-1. The authors acknowledge support by the state of Baden-Württemberg through bwHPC.}

\usepackage[square,numbers]{natbib}
\EventAcronym{ESA}
\EventYear{2024}
\ArticleNo{XX}
\ccsdesc[500]{Mathematics of computing~Graph algorithms}
\ccsdesc[300]{Mathematics of computing~Paths and connectivity problems} %
\keywords{edge orientation, pseudoarboricity, algorithm engineering} %

\usepackage{xspace}
\usepackage{wrapfig}
\usepackage{graphics}
\usepackage{todonotes}
\usepackage{booktabs}
\usepackage{csquotes}

\usepackage{afterpage}

\newcommand{\TwoApproxString}{TwoA}
\newcommand{\ALGODFS}{\textit{Dfs}\xspace}
\newcommand{\ALGODFSNAIVE}{\textit{DfsNaive}\xspace}
\newcommand{\ALGODFSNAIVESHARED}{\textit{DfsNaiveShared}\xspace}
\newcommand{\ALGODFSNAIVEWITHOUTSHARED}{\textit{DfsNaiveWithoutShared}\xspace}
\newcommand{\ALGOEAGERPATHSEARCHPLUSDFS}{\textit{EPS+Dfs}\xspace}
\newcommand{\ALGOEAGERPATHSEARCHPLUSBFS}{\textit{EPS+Bfs}\xspace}

\newcommand{\ALGOBATCHEDBFS}{\textit{Bfs}\xspace}
\newcommand{\ALGOFASTIMPROVE}{\textit{FastImprove}\xspace}
\newcommand{\ALGOEAGERPATHSEARCH}{\textit{EPS}\xspace}
\newcommand{\ALGOTWOAPPROX}{\textit{\TwoApproxString}\xspace}
\newcommand{\ALGOTWOAPPROXCONDITIONAL}{\textit{TwoADens}\xspace}
\newcommand{\ALGOFINALNAME}{\textit{RapidPathOrientation}\xspace}

\newcommand{\COMPETITORKOWALIK}{\textit{Kowalik}\xspace}
\newcommand{\COMPETITORKOWALIKTwoApprox}{\textit{\TwoApproxString+Kowalik}\xspace}
\newcommand{\COMPETITORGP}{\textit{G\&P}\xspace}
\newcommand{\COMPETITORGPTwoApprox}{\textit{\TwoApproxString+G\&P}\xspace}

\author{Henrik Reinstädtler}{Heidelberg University, Germany }{henrik.reinstaedtler@informatik.uni-heidelberg.de}{https://orcid.org/0009-0003-4245-0966}{}%
\author{Christian Schulz}{Heidelberg University, Germany }{christian.schulz@informatik.uni-heidelberg.de}{https://orcid.org/0000-0002-2823-3506}{}%
\author{Bora Uçar}{CNRS and LIP, ENS de Lyon, France \and UMR5668 (CNRS, ENS de Lyon, Inria, UCBL1) France \and 
\url{http://perso.ens-lyon.fr/bora.ucar}}{bora.ucar@ens-lyon.fr}{https://orcid.org/0000-0002-4960-3545}{}

\authorrunning{H. Reinstädtler, C. Schulz and B. Uçar}
\begin{document}

\title{Engineering Edge Orientation Algorithms}

\date{}

\maketitle

\begin{abstract} \small%

Given an undirected graph $G$, the edge orientation problem asks for assigning a direction to each edge to convert $G$ into a directed graph.  The aim is to minimize the maximum out degree of  a vertex in the resulting directed graph.  This problem, which is solvable in polynomial time, arises in many applications.  An ongoing challenge in edge orientation algorithms is their scalability, particularly in handling large-scale networks with millions or billions of edges efficiently.  We propose a novel algorithmic framework based on finding and manipulating simple paths to face this challenge.  Our framework is based on an existing algorithm and allows many algorithmic choices.  By carefully exploring these choices and engineering the underlying algorithms, we obtain an implementation which is more efficient and scalable than the current state-of-the-art.  Our experiments demonstrate significant performance improvements compared to state-of-the-art solvers. On average our algorithm is 6.59 times faster when compared to the state-of-the-art.  
\end{abstract}

\pagenumbering{arabic}

\section{Introduction}
\label{sec:introduction}
Graphs and networks play a vital role in our connected society for modelling and understanding complex problems. A graph consists of a set of vertices connected by edges, which may be directed or undirected, depending on the specific modeling requirements.
For some applications, such as stabilizing telecommunication networks~\cite{venkateswaran2004minimizing}, it is necessary to orient each edge of an undirected graph, thereby converting it into a directed graph. 
In telecommunication networks a lower number of outgoing edges equates to a higher fault tolerance, 
as not too many connections would be affected by a fault in one of the connection hubs modelled as vertex.
One frequently used quality metric for an orientation  is the maximum out-degree of a vertex.
Given an undirected graph $G$, the \emph{edge orientation problem} asks for an orientation of $G$ in which the maximum out-degree of a vertex is minimized. 

The edge orientation problem has a wide range of applications~\cite{georgakopoulos2007max}. 
Besides stabilizing telecommunication networks, other applications  include  storing optimal graphs~\cite{aichholzer1995optimal} or analysis of structural rigidity~\cite{doi:10.1137/0401025}.
In map labeling~\cite{kakoulis1998multiple} dense areas of maps can only have one label.
In order to formalize this concept, one related task is to identify the densest sub-graph, which is the set of vertices, that has the highest edge to vertex ratio.
This task is called the max-density task and has further applications in bioinformatics, web analysis and scheduling~\cite{georgakopoulos2007max}. For more applications and their details we refer the reader to Georgakopoulos and Politopoulos~\cite{georgakopoulos2007max}.

There are different approaches to solve the edge orientation problem exactly.
One of the simplest algorithms is by Venkateswaran~\cite{venkateswaran2004minimizing} which solves the problem  in $\mathcal{O}(m^2)$ time.
The core idea of this algorithm is to repeatedly find paths with a breadth-first-search from a vertex having high out-degree to a vertex having low out-degree.
Once such a path is found, the orientation of each of the edges on the path is reversed. 
Thus by this operation the degree of the high out-degree vertex is reduced and conversely, the degree of the low-degree vertex is increased. 
Other  approaches solve the problem by using a flow based formulation~\cite{asahiro2007graph,kowalik2006approximation}. 
Kowalik~\cite{kowalik2006approximation} gives an approximation scheme for this problem.
These approaches analyse the theoretical aspects, without in depth practical study.
The current best complexity bounds and the only implementation in Java of an exact version of Kowalik's flow based solution are presented by Blumenstock~\cite{doi:10.1137/1.9781611974317.10}. %
However, an ongoing challenge in edge orientation algorithms is their scalability, particularly in handling large-scale networks with millions or billions of edges efficiently. 
As real-world networks continue to expand in size and complexity, there is a need for algorithms capable of scaling effectively to such massive datasets. %

\textit{Contribution.} We introduce a novel framework of algorithms inspired by Venkateswaran work~\cite{venkateswaran2004minimizing} of improving paths to tackle the edge orientation problem. 
Our experimental results show the advantages of this approach over conventional flow-based formulations. Additionally, we offer an alternative proof of correctness for this framework. %
In addition to a verbatim implementation of the original algorithm, we provide an accelerated version in which we engineer all components of the algorithm. This includes efficient pruning of the search space, batch search of improving paths and better initialization routines that are used as starting point of the algorithm.
Experiments show that our algorithm can scale to huge instances and that the performance improvements of our algorithm over the state-of-the-art are very large.
 For example our fastest algorithm is on average 6.59 times faster than our (faster) C++ reimplementation of the state-of-the-art~solver. %

This paper is organized as follows: We first introduce basic concepts and provide an extensive review of related work in Section~\ref{sec:preliminaries} and \ref{subsec:related_work}. Section~\ref{sec:impr} provides an alternative proof of correctness using a flow-based formulation on a bipartite representation of the graph for the algorithm by Venkateswaran. We then proceed to discuss several speed-up techniques, including eager depth-first search, and explore various variations of breadth-first and depth-first search in Section~\ref{sec:impr}. These new approaches are extensively benchmarked in Section~\ref{sec:Experimental Evaluation}, followed by our conclusions in Section~\ref{sec:conclusion}.

\section{Preliminaries}
\label{sec:preliminaries}
\subparagraph{Orientation Problem.} An undirected graph $G=(V,E)$ consists of a vertex set $V$ containing $n$ vertices and an edge set $E \subseteq \binom{V}{2}$ of $m$ edges.
An orientation $O$ assigns %
a direction to each edge in $G$ and results in a directed graph.
We denote the direction of an edge by $O(e)=u\to v$, if $e$ is oriented from $u$ to $v$.
For an edge $e=\{u,v\}$ oriented as $u\to v$, we say that $e$ is an outgoing edge of $u$ and incoming edge of $v$.
The out-degree $d(O,v)$ of a vertex $v$ in an orientation $O$ is defined as the number of outgoing edges of $v$, that is, $d(O,v)=|\{u: v\to u \in O\}|$.
If the orientation is clear from the context, we write $d(v)$.
Similarly, the in degree of a vertex is the number of its incoming edges.
We call the vertices with the largest  out-degree in an orientation \emph{peak vertices}.
A path in an orientation is a finite sequence of vertices $u_1,\dots,u_n$, where there is an edge oriented $u_{i}\to u_{i+1}$. We call a path simple if no vertex is contained twice in the path.
In an orientation, changing the orientation of an edge from $v\to u$ to $u\to v$ is called flipping.
A path can be flipped by flipping all edges contained in it once.
For a given graph $G$, the \emph{edge orientation problem} asks for an orientation $O$ such that $d^\star=\max_{v\in V}{d(O,v)}$ is minimized.

\subparagraph{(Pseudo-)arboricity.}
A forest is a collection of edges, where each vertex is only connected by one path.
The edges of an undirected graph can be partitioned into disjoint forests and the minimum number of forests is known as the \emph{arboricity}.
A more relaxed version of this problem is to decompose the graph into pseudoforests, where in every connected component there can be at most one cycle.
The minimum number of pseudoforests partitioning the edges is called \emph{pseudoarboricity}. The pseudoarboricity is known to be equivalent to the maximum out-degree of an optimal edge orientation~\cite{venkateswaran2004minimizing}.
The average density of a graph is defined as the ratio of edges to vertices.
A sub-graph contains only edges between  a subset of vertices. 
 The maximum average density of any sub-graph is closely related to the pseudoarboricity.
As shown by Picard and Queyranne~\cite{picard1982network} and Venkateswaran~\cite{venkateswaran2004minimizing} the pseudoarboricity  is equal the ceiling of the maximum average density of any  sub-graph.
Picard and Queyranne~\cite{picard1982network} show the arboricity is equal to either the pseudoarboricity \hbox{or the pseudoarboricity plus one.}
\subparagraph{Integral Flows.}
Given a directed graph $G=(V,E)$, a source vertex $s\in V$, a target vertex $t\in V$, and an edge capacity function $c:E\to \mathbb{N}$, a  flow is a function $f:E\to \mathbb{N}$ that satisfies two conditions: (i)
the flow does not exceed the capacity in any edge; (ii) and the inflow in every vertex equals the outflow, except for $s$ and $t$, which have respectively positive out and in flows.
Given a flow, one can define a residual capacity for each edge, which is equal to edge's original capacity minus the current flow.
A residual network for a given flow consists of all edges with a positive residual capacity.
If the flow on an edge is equal its capacity, the edge is called saturated. 
When changing the capacity of an edge, an edge  can become under or over-saturated and the flow needs to be updated.
\subparagraph{Bipartite $b$-Matching.} 
In a bipartite graph the vertex set can be partitioned in disjoint sets $S$, $T$ such that all edges contain one vertex from $S$ and $T$ each.
A matching $M$ is a set of edges no two of which share a vertex.
A maximum matching is a matching with the largest number of edges.
For a given positive integer $b(v)$ for each vertex $v$, 
the $b$-matching problem asks for a set $F$ of edges with the largest cardinality such that 
the vertex $v$ is included in at most $b(v)$ in $F$.
A $b$-matching in a bipartite graph $(S\cup T, E)$ can be found by modeling it as a flow problem as follows.
First, add a source vertex $s$ and a sink $t$ to the graph. Then add edges between $s$ every vertex in $v$ in $S$ with capacity $b(v)$ as well as edges between each vertex $w$ in $T$ with capacity $b(w)$ from $w$ to $t$.
 The original edges are assigned unit capacity.
The saturated edges of an integral max-flow from $s$ to $t$ are the edges in an~optimal~$b$-matching.

\section{Related Work}
\label{subsec:related_work}
\subparagraph{Edge Orientation Task and Pseudoarboricity.}
\begin{algorithm}[b!]
  \begin{algorithmic}[1]
    \PROCEDURE{Venkateswaran}{$G=(V,E)$}
    \STATE $O \gets \textrm{an arbitrary orientation of } G$
    \STATE $k\gets\max_{v\in V}d(O,v)$
    \STATE $S\gets\{ v\in V \mid d(O,v)= k\}$
    \STATE $T \gets\{ v\in V \mid d(O,v)\leq k-2\}$
    \WHILE{BFS finds path $P$=$s, \ldots, t$ from $S$ to $T$ in $O$}
    \STATE Flip $P$ in $O$ %
    \STATE Remove $s$ from $S$
    \STATE Remove $t$ from $T$ if $d(O,t)=k-1$
    \IF{$S$ empty}
    \STATE $k\gets k-1$
    \STATE $S\gets\{ v\in V \mid d(O,v)= k\}$
    \STATE $T \gets\{ v\in V \mid d(O,v)\leq k-2\}$
    \ENDIF
    \ENDWHILE
    \Return $k$
    \ENDPROCEDURE
  \end{algorithmic}
  \caption{Algorithm by Venkateswaran~\cite{venkateswaran2004minimizing}}
  \label{alg:venkateswaran}
\end{algorithm}
There has been a wide variety of approaches to solve the edge orientation task and the identical pseudoarboricity task.
One algorithm is by Venkateswaran~\cite{venkateswaran2004minimizing} and underlies the algorithms proposed in this work.
Venkateswaran gives an algorithm for computing an extremal orientation to minimize the maximum in degree. 
The algorithm can be easily translated to the out-degree setting, as shown 
in Appendix Algorithm~\ref{alg:venkateswaran}.
After arbitrary initializing an orientation, the algorithm starts to search for improvements by finding paths between the set of vertices~$S$ with max out-degree $k$ and the set of vertices~$T$ with out-degree strictly lower than $k-1$.
If $S$ is empty, then $k$ is reduced  by one, and  the sets $S$ and $T$ are reinitialized.
 If no path is found, the current $k$ is returned as optimal.
The correctness of this algorithm follows from the density of the sub-graph induced by the vertices visited by the failing BFS starting from $S$. %
 These have a degree of at least $k-1$ and there is at least one with degree $k$, totaling an average density greater than $k-1$, leading to a pseudoarboricity of $k$ by the sub-graph density argument.
It is proven that the running time of this algorithm is $\mathcal{O}(m^2)$, since each path can be found in $\mathcal{O}(m)$ and the number of improvements can be bounded by $m$ as well.
The argument is that the total number of paths for one vertex is bounded by its out-degree, which again is bounded by \hbox{the number of edges in total.}

A $2$-approximation of the pseudoarboricity or the maximum average density  can be found in linear time \cite{charikar2000greedy,georgakopoulos2007max} by repeatedly deleting min degree vertices.  %
Georgakopoulos and Politopoulos~\cite{georgakopoulos2007max} provide an algorithm for finding the densest subset in a more general setting of set systems, improving on ideas by Goldberg~\cite{goldberg1984finding} by pruning the graph during their binary search scheme.
The algorithm can be used to compute the max out-degree, but does not find a suitable~orientation.

Asahiro~\etal~\cite{asahiro2007graph} give a solution with running time $\mathcal{O}(m^{3/2}\log{d^\star})$ based on flows for the problem and offer more results for approximating the related weighted orientation problem.
Kowalik~\cite{kowalik2006approximation} give an approximation scheme using flows that can be used to calculate exact solutions as well. They construct a virtual graph with a source and a sink vertex. Given an orientation $O$ and a test value $d'$ they add an arc between the source vertex and vertices with higher than $d'$ out-degree with capacity $d(O,v)-d'$.  Each vertex with lower than $d'$ out-degree is connected to the sink with capacity $d'-d(O,v)$. For each oriented edge $u\to v$, there is an arc $v$ to $u$ with capacity $1$. The flow is then computed and all edges with a corresponding exhausted arc are flipped. If there is an arc from the source without flow, the test failed and the maximum out-degree must be higher.  With a binary search the optimal value $d^\star$ and a suitable orientation can be found in $\log d^\star$ steps.

Blumenstock~\cite{doi:10.1137/1.9781611974317.10} presents bounds for the aforementioned flow based solution using Dinic's algorithm and almost unit capacity networks.
 The best general worst case bound for the problem is $\mathcal{O}(m^{3/2}\sqrt{\log{\log{d^\star}}})$.
Moreover, Blumenstock~\cite{doi:10.1137/1.9781611974317.10} also presents the first practical evaluation of algorithms for the pseudoarboricity problem by implementing  the flow based algorithms by Kowalik~\cite{kowalik2006approximation}  and Georgakopoulos and Politopoulos~\cite{georgakopoulos2007max}. %
Experiments shows that  Georgakopoulos and Politopoulos's method using unit capacity networks is faster than other flow based solutions, including Kowalik's algorithm. 
However, subsequent analysis revealed a subtle implementation error in the implementation of the algorithm which caused it to output incorrect results in rare cases. Once corrected, the performance advantage of this method is no longer apparent.
We use both of these algorithms as the state-of-the-art in our experiments. %

\vspace*{-.5cm}
\subparagraph{Bipartite Matching and Network Flows.}
Finding a $b$-matching can be done by transforming the problem into an uncapacitated matching problem as described by Gabow~\cite{gabow1983efficient}: vertices (and their respective edges) are replicated according to their capacity.
Bipartite matching is a well-studied and understood part of computer science.
The best known worst-case algorithm for solving bipartite matching is Hopcroft-Karp~\cite{hoka:73} with a complexity of $\mathcal{O}(m\cdot \sqrt{n})$. It can be seen as a specialization of Dinic's algorithm for general flows.
For graphs with capacity $1$ per vertex, like in the bipartite matching case,  the complexity of Dinic's algorithm is $\mathcal{O}(n^{1/2}m)$ according to Tarjan and Even~\cite{even1975network} matching the bound from Hopcroft-Karp.
For general flows Dinic's algorithm has a complexity of $\mathcal{O}(n^2m)$ and better alternatives like Goldberg and Tarjans algorithm~\cite{goldberg1988new} are available in practice.%

\ifFullPaper
\subparagraph{Related Problems.} 
The edge orientation task can be also done on dynamic graphs.
Bender~\etal~\cite{bender2021incremental} show an amortized constant time algorithm for maintaining an out-degree of up to $3$ in an incremental forest, their use case is Cuckoo hashing.
Most recent, Borowitz~\etal~\cite{DBLP:journals/corr/abs-2301-06968} developed a promising approach for orienting fully dynamic graphs.
Another related task is the coloring task, which is related via the arboricity since the arboricity can be used to determine upper limits on the chromatic number. Christiansen~\etal~\cite{dynamiccoloringsparsegraphs} recently presented an algorithm using $\mathcal{O}(\alpha^2)$ colors for dynamic graph coloring on sparse graphs with~$\alpha$~arboricity.
\else
\fi

\section{Edge Orientation Framework and Engineering Techniques}\label{sec:impr}
\begin{wrapfigure}{r}{0.5\linewidth}
  \centering
  \includegraphics[width=6cm,page=6]{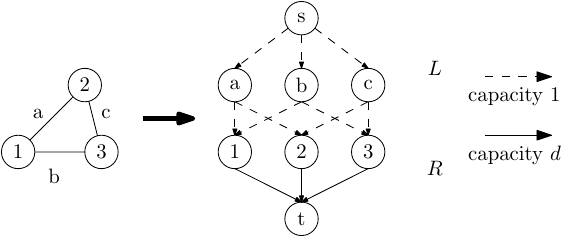}
  \caption[]{Example bipartite repr.~transformation.}
  \label{fig:setup}
  \vspace*{-.5cm}
\end{wrapfigure}
We give an interpretation of the working of Algorithm~\ref{alg:venkateswaran} using maximum flows and $b$-matchings in bipartite graphs.
This interpretation allows us to identify key aspects for engineering to improve the practical running time of the algorithm.%
For a given graph $G=(V,E)$, we construct a bipartite graph $B_G=(L\cup R, E_B)$ with $|E|+|V|$ vertices, and $2|E|$ edges.
Roughly speaking, $L$ will represent the edges of the original graph and $R$ will represent the vertices of the original graph.
For each edge $e=\{u,v\}\in E$, we  have a vertex $\ell(e)\in L$.
For each vertex $v\in V$, we have a vertex $r(v)\in R$.
For each $e=\{u,v\}\in E$, we have two edges in $E_B$: one between $\ell(e)$ and $r(u)$
and another between $\ell(e)$ and $r(v)$.
In this bipartite graph, an edge orientation can be described by an assignment of each $L$ vertex to one of its two neighbors in $R$.
Here, each $L$-vertex is assigned to a unique $R$-vertex, and each $R$-vertex can have multiple $R$-vertices assigned to it.
The edge orientation problem can thus be solved by finding an assignment of $L$ vertices to $R$ vertices
which minimizes the maximum number of $L$ vertices assigned to an $R$-vertex.
This can be solved \hbox{by a network flow formulation.}

Our network flow formulation starts with $B_G$ and
adds two special vertices $s$ and $t$ to $B_G$. The vertex $s$ is connected to all $L$ vertices, and all $R$-vertices are connected to the vertex~$t$.
Then, a capacity of $1$  is attributed to
all edges between $s$ and $L$-vertices, and also to all edges of $B_G$; that is from a vertex $\ell(e)$ to the vertices $r(u)$ and $r(v)$ for $e=\{u,v\}\in E$.
If we attribute a capacity of $d$ to edges $(r,t)$ and  have a feasible flow with a total flow value of $|E|$ (out of $s$), then all edges can be oriented while having an out-degree of at most $d$ (see below).
An example for a simple graph $G$ of three vertices and three edges, the flow graph obtained by adding $s$ and $t$ to $B_G$ can be found in Figure~\ref{fig:setup}.

Due to the capacity of $1$ for each $(s,\ell)$ edge, every edge in the original graph will be oriented in only one direction. 
For an edge $e=\{u,v\}$ in the original graph, the edge is oriented from $u\to v$, if in $B_G$ there is a flow from $\ell(e)$ to $r(u)$ and $v \to u$, if there is a flow from $\ell(e)$ to $r(v)$.
Each edge is assigned a direction, since we search for a flow in which all $|E|$ edges in the flow graph of the form $(s, \ell)$ are saturated.
Clearly, the smallest value of $d$ allowing this will be the optimal $d^\star$.
When a smaller value is tested and the out flow in $s$ is not equal $|E|$, some edges are not assigned a direction.
Since this is a classical capacitated network flow problem with integer capacities, an integer solution can be found
with augmenting path-based algorithms~\cite[Theorem 26.10]{clrs:09}.
A binary search for the optimal~$d$ will have in the worst case $\log d_G$ steps, where $ d_G$ is the maximum degree of a vertex in $G$.
Instead of running a flow algorithm from scratch, the results of previous searches can be reused.
When increasing the value of $d$, the previous flow can be used as a starting point. 
If we decrease the test value,  the flow in oversaturated edges starting in $R$ must be balanced first by reducing the flow along a path starting in $s$ to $t$.
 Afterwards any  augmenting path needs to be found.

\textbf{Correspondence to Venkateswaran's Algorithm.}
Given an arbitrary orientation  for the out-degree, we can construct the flow in the bipartite representation as follows:
For each  edge $e = \{u,v\}$, oriented $u\to v$ we set the flow from $\ell(e)$ to $r(u)$ to $1$ and $\ell(e)$ to $r(v)$ to $0$.
In the residual network there is a residual capacity of $1$ between $r(v)$ and $\ell(e)$.
The obtained flow network is saturated in all edges $(s,l), l\in L$, since each edge is assigned a direction and therefore has a flow from  $\ell(e)$ to either $u$ or $v$. %
The initial test value for the max out-degree is set to the maximum out-degree of the orientation and repeatedly decreased,
until some edges from $s$ to $L$ become unsaturated. %

In each decreasing step there is an oversaturated edge $(r(v),t)$, that needs to be decreased by one, at most for every $v$ and along to a path to $s$ the flow needs to be reduced by one. 
This path has length 2, since one arbitrary edge $(\ell(e),r(v))$ with flow greater zero will be chosen and its flow as well as the flow on the single incoming edge $(s,\ell(e))$ reduced.
Afterwards an augmenting path needs to be found.
 This path must include $\ell(e)$ as second vertex since all other edges $(s,L)$ are saturated.
Instead of first reducing and then finding an augmenting path, we can combine both steps.
 In other words, the flow algorithm needs to find distinct paths in the residual network between all vertices in $R$ with oversaturated edges to $t$ and vertices in $R$, that have edges towards edges with enough residual capacity. %
This can be  done using breadth-first-search algorithms.
The  vertices in $R$ with oversaturated edges are exactly the vertices in $S$,  since we decreased the test value by one. 
Similarly, the last vertices before $t$ in augmenting paths need to be vertices $r(v)\in R$ with residual capacity in $(r(v),t)$ and enough unsaturated incoming edges.
These vertices are exactly those vertices $T$ in Venkateswaran algorithm.
 Due to the bipartite structure paths must include the vertices in~$L$.
The edge vertices $L$ in a residual network of $B_G$ can be only traversed in one direction by the BFS, exactly like the BFS in Venkateswaran.
The only difference, compared to Venkateswaran, is that in our network, the path is twice as long, since each edge in $G$ consists out of two subedges in $B_G$.
When in our model a too low test value is set, augmenting paths  cease to exist.

\subsection{The Proposed Framework}
\begin{wrapfigure}{r}{0.55\linewidth}
\vspace*{-.95cm}
\begin{minipage}{\linewidth}
\begin{algorithm}[H]
  \begin{algorithmic}[1]
    \PROCEDURE{ExhaustiveSearch}{$G=(V,E),O$}
    \STATE \Repeat\label{alg:whilepass}
    \FOR{$v\in V$ with $d(v,O)= \max_{w\in V} d(O,w)$}
    \IF{$P=\textsc{FindPath}$($G,O,v$) exists}
    \STATE Flip $P$
    \ENDIF
    \ENDFOR
    \Until{no path has been flipped in last iteration}
    \RETURN $O$
    \ENDPROCEDURE
  \end{algorithmic}
  \caption{Orientation by Exhaustive Search.}
  \label{alg:overview}
\end{algorithm}
\end{minipage}
\end{wrapfigure}
We now present our techniques and framework to improve the performance of Venkateswaran's algorithm in practice.
We introduce data structures and algorithmic choices inspired by the maximum cardinality matching problems~\cite{duff:81a,duku:12,hoka:73,pofa:90}. 
Engineering techniques include a simple orientation improving algorithm with complexity $\mathcal{O}(m)$ that yields a much better starting point for the path finding algorithm in practice.  Moreover, we enhance the efficiency of path searches by either reusing information or batching the searches and exploring paths in reverse order. %

Our general algorithm framework is shown in Algorithm~\ref{alg:overview}. It is an adapted version of the algorithm by Venkateswaran.
Instead of constructing the sets $S$ and $T$ explicitly we iterate over all vertices that currently have the largest out-degree and try to find an improving path, which is a path in the orientation from one of these vertex to a vertex with lower out-degree.
If such a path is found, it is flipped, resulting in a lower out-degree for the start vertex.
 The algorithm continues this process until no such path exists.
 The correctness of this general algorithm follows from Venkateswaran's algorithm.
 It has an equivalent termination criterion as the original algorithm, since when it finishes when no further improving path starting at peak vertices can be found.
 Instead of searching from all peak vertices at the same time, we search for every peak vertex separately.
 Like the algorithm by Venkateswaran this algorithm has a time complexity of $\mathcal{O}(m^2)$ since we need to find at most $m$ improvements by the argument of  Venkateswaran~\cite{venkateswaran2004minimizing}.
 Furthermore, each improvement can be found with one DFS or BFS in $\mathcal{O}(m)$, yielding its total complexity.
\subsection{Engineering Techniques}\label{sec:engineering}
 We now discuss how we can improve the techniques to compute orientations.
 First, we discuss the linear $2$-approximation data reduction and come up with a fast initialization algorithm.
 Secondly, we review different ways of finding improving paths and general ideas to optimize this search.

\subsubsection{Two Approximation Data Reduction}\label{sec:2approx} 
 The \emph{linear} $2$-approximation proposed by Charikar~\cite{charikar2000greedy} can be used as data reduction. It computes  a $2$--approximation $d_{approx}$ using a bucket priority queue and vertices with degree $\frac{d_{approx}}{2}$ or less can be removed safely due to the densest sub-graph argument. In other words, since the $d_{approx}$ is a $2$--approximation, the optimum outdegree must be at least $\frac{d_{approx}}{2}$. Vertices with a degree that is smaller or equal to this value can therefore be removed iteratively. The edges of these vertices are oriented outwards.  
The approximation needs to process every edge of the graph at least once and maintaining the needed priority queue is not negligible.
 Therefore, running it always as a preprocessing step can be costly, if the resulting guess is small and does not remove much of the graph.
 Instead, we propose to run the $2$-approximation conditionally on the density of a graph $\rho=\frac{m}{n}$.
 If the density is smaller than some parameter $c$, we directly skip to find improving paths, instead of wasting time to compute a low approximation.
 In experiments we  tune this parameter.

\subsubsection{Fast Initialization} 
\begin{wrapfigure}{r}{0.48\linewidth}
  \begin{minipage}{\linewidth}
    \vspace*{-1cm}
\begin{algorithm}[H]
  \begin{algorithmic}[1]
    \PROCEDURE{FastImprove}{$G=(V,E),O$}
    \FOR{$v\in V$}
    \FOR{$e=v\to u\in O$}
    \IF{$ d(O,u)< d(O,v)-1$}
    \STATE Flip($e$)
    \ENDIF
    \ENDFOR
    \FOR{$e=u\to v\in O$}
    \IF{$ d(O,u)-1> d(O,v)$}
    \STATE Flip($e$)
    \ENDIF
    \ENDFOR
    \ENDFOR
    \RETURN $O$
    \ENDPROCEDURE
  \end{algorithmic}
  \caption{FastImprove Algorithm}
  \label{alg:resort}
\end{algorithm}
\end{minipage}
\end{wrapfigure}
\noindent Venkateswaran's algorithm starts with an arbitrary orientation and thus has worst-case complexity $O(m^2)$. 
However, it is evident that starting with a good orientation can lead to a faster algorithm compared to an arbitrary initialization.  
The \textsc{FastImprove}-Algorithm in Algorithm~\ref{alg:resort} makes one pass over an arbitrary orientation and flips an edge $v\to u$, if such a flipping improves the largest out-degree of $v$ and $u$. 
Since only direct improvements are found, the algorithm does not find the optimal solution. However, it can improve a given orientation significantly.
In our implementation, vertices are represented by their IDs which are numbers in the range $0,\ldots,n-1$. 
Our algorithm initializes an orientation with $u\to v$ for~$u>v$ and then applies the \ALGOFASTIMPROVE algorithm to quickly improve it. %
The complexity of one pass is $O(m)$.
In the experimental evaluation in Section~\ref{sec:Experimental Evaluation}, we show that this simple algorithm can tremendously speed up computation time of the overall algorithm.

\subsubsection{Path Finding Algorithms} \label{sec:path:find:alg}
There are two basic ways to find paths in a directed graph: either breadth or depth first search.
Breadth first search works by using a queue and adding neighboring vertices following first in first out order.
Depth first search is usually implemented using a stack and a visited array.
The visited array stores whether a vertex has been visited before, while the stack keeps track of the current path.
The next not visited neighboring vertex is always added to the stack and marked as visited, the neighbors are explored recursively using the stack, \ie depth first.
 Once all neighbors are explored the previous vertex of the stack is (re-)visited and its remaining neighbors are visited.

\textbf{Batched BFS.}
The batched BFS approach finds improving paths similar to the original idea by Venkateswaran of an adapted BFS.
It does not fit the previous described framework in Algorithm~\ref{alg:overview}.
Instead of starting a BFS from every vertex, we start the BFS by putting  all peak vertices  in the queue and flip a path once we have found an improving path by breadth first search. 
We continue the search until all vertices have been reached or, in case of success, we found an improvement for all peak vertices.
The search is continued until for none of the vertices we find an improving path.
This yields a correct result, since we inserted all peak vertices at the start and thus will find a shortest improving path, if there exists one.
 However, we will not always find a path for all peak vertices at once.

\textbf{DFS.}
We now look at ways to improve depth-first-search.
Our ideas include checking neighboring vertices eagerly, ensuring independent paths by visiting peak vertices only as starting vertices of paths, reusing the visited array and eagerly ordering path searches.

\textit{Early Check.}
The first improvement is to check the out-degree of all neighbors before continuing exploring them in the recursion.
 This increases the overall complexity by a constant factor, but in practice speed up the exploration massively by preventing unneeded recursion. %

\textit{DFS with Independent Paths.} We propose for the DFS to not continue traversal via other peak vertices while searching improving paths. There is no added benefit in adding a peak vertex to a path during search as each peak vertex needs its own improving path. 
More precisely, they have to be distinct, since flipping one of the paths would change the orientation of the joint part of the two paths, effectively stopping the  second path from being valid. %
 Hence, we need to find independent paths from both of the vertices regardless.

\textit{DFS with Shared Visited Array.}
A DFS can reuse information stored in the visited array for multiple consecutive searches for improving paths. %
Here, we reset the visited array used in a classical DFS only after one pass over all peak vertices and not after every path search. 
Thus, we enhance efficiency by retaining the visited array's state across multiple consecutive searches to identify improving paths, rather than resetting it after each search. This method strategically excludes previously explored and non-improving sub-graphs from subsequent searches, significantly reducing computational redundancy. By maintaining the visited array across searches, we ensure that once a path from a peak vertex is improved, subsequent searches do not redundantly explore the same paths or sub-graphs \hbox{already deemed~non-improving.}
\begin{wrapfigure}{r}{0.41\linewidth}
   \includegraphics[page=3]{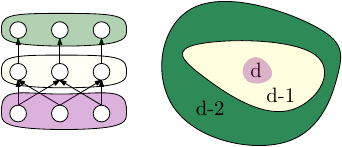}
   \caption{\textbf{Left:} Layered example orientation in a graph with nine edges and vertices. \textbf{Right:} Visualization of the onion-like structure of the orientation problem.%
   }
   \label{fig:mdfs}
 \end{wrapfigure}
Figure~\ref{fig:mdfs} (Left) gives a simple example orientation, where not resetting the visited array is helpful:
 If the first path from the bottom left to the top left vertex is found and flipped, we do not want the search starting in the middle vertex to traverse via left tree that was already explored during the first DFS.
 Moreover, this example orientation shows, why this DFS approach is better than the BFS described in the earlier section. The BFS would find at most two improving paths, because it eagerly finds predecessors.
 It would assign the middle vertices (yellow) only two distinct predecessors at most due to the graph structure, leaving one peak vertex (purple) without successors and leading to two paths for the bottom left vertex.
 However, we are only interested in one path per peak vertex.
 Thus, this leads to more path searches than necessary.

\textit{Eager Path Search.}
The \ALGOFASTIMPROVE algorithm only uses direct improvements and  reduces their number, leaving longer improving paths to be discovered.
Therefore, our problem has an onion-like structure, multiple layers of vertices with the same out-degree are surrounding some core peak vertices after the initialization. %
By our classical approach we would find improvements from core to the lower degree outer layers and slowly growing the number of peak vertices.
This leads to prolonged paths, as there will be multiple paths needed to be found while decreasing the  out-degree of a peak vertex by one at a time. %
 An example for this is shown in Figure~\ref{fig:mdfs} (Right), the vertices with out-degree $d$ will need to be improved multiple times via the vertices with out-degree $d-1$.
 Instead of slowly searching consecutively, we propose to first find paths  for the outer layers of the problem. We define $i$ as the number of layers we reduce eagerly in the reverse order of their out-degree,\ie $d-i$, $d-i+1,$ and so on. %
 Moreover,
 if only fewer vertices than some parameter $c$ have an out-degree of $d-i+k$  in the orientation, we propose to repeatedly search up to $k$ improvements with a DFS.
 Intuitively, this reduces the number of times the algorithm needs to collect these vertices and find paths. %

  As both ideas are combined with the idea of the reused visited array, this method does not return an exact solution always: 
Suppose $d$ is the current max out-degree, then it can occur, that there is no improvement for $d-1$, but there would exist some path from $d$ out-degree vertices to $d-2$ out-degree vertices through $d-1$ vertices.
Since the $d-1$ vertices are checked first, there is no improving path found and the visited information is kept leading to all $d-1$ vertices marked visited.
The searches started in $d$ can not traverse over these vertices and the search is unsuccessful.
   Therefore, it requires to run either a DFS or a BFS afterwards.
We propose to decide based on the maximum size of the layers and the resulting max out-degree whether to run a batched BFS or DFS.

Moreover, the choice of $i$ and $c$ is crucial, in Section~\ref{sec:Experimental Evaluation} we test static values for $c$ and $i$.
Additionally, we test for $i$  a dynamic value of $\sqrt{\max d -\rho}$ with $\rho = \frac{m}{n}$ as average density of~the~graph and $\max d$ being the out-degree of the starting orientation.
The maximum out-degree of the starting orientation $\max d$ and $\rho$ are the natural lower and upper bounds for this problem. 
 We chose the square root as a natural damping function in order to not explore too many layers eagerly.

\section{Experimental Evaluation}
\label{sec:Experimental Evaluation}

\subparagraph{Methodology.}
We implemented our algorithms and data structures in \cpp\textsf{20}.
We compiled our program and all competitors using  \textsf{g}\texttt{++-}\textsf{12.1} with full optimization turned on (\texttt{-O3} flag).
In our experiments, we use two types of machines provided by a cluster for our experiments.
 Both machines are equipped with  $20$-core Intel~Xeon~Gold~6230  processors running at $2.10$ GHz and having a cache of 27.5 MB. Machine type A has two sockets and 96 GB of RAM, machine type B has four sockets with 3 TB of RAM.
For the  five graphs with more than 1 billion edges we use the machine type B.
We run each algorithm on each
 instance five times and use the arithmetic mean of the running time of these independent runs in the experiments.
Up to four experiments were run in parallel on machine type A and 6 on machine type B. The order of the experiments was random. %
The experiments that did not finish within 5 hours were only repeated once.
 The running times reported include the setup of data structure needed by the algorithms, but does not include the initial loading of the graphs from the hard drive.
For comparisons, we use performance profiles as proposed by Dolan and Mor{\'e}~\cite{performanceprofiles}. 
We plot which fraction of instances is solved by an algorithm within $\tau t_{opt}$, for $ \tau\geq 1$ and $t_{opt}$ being the best running time reported by any algorithm on a given instance. 
A~higher fraction at a lower $\tau$ means that more instances are solved within this $\tau$, implicating a better performance.

        \subparagraph{Instances.} We use graphs from the SuiteSparse Sparse Matrix Collection~\cite{floridasparsematrix} with more than~1~million vertices for benchmarking. The set contains 67  graphs from a wide range of applications. Statistics on the instances, including the number of edges, the number of vertices, the minimum and the maximum degree of a vertex, and the number of connected components, are shown in Table~\ref{tab:instancedetails} in the Appendix.
 The largest graph in our set has more than five billion edges. %
Based on the average density of~the~graphs, we chose nine representative instances (\textbf{bold} in Table~\ref{tab:instancedetails}) for 
evaluating different parameters for the proposed algorithms and 
 running preliminary experiments faster. 
These representative instances are called \textbf{test set} in the following and are not included in the geometric means and performance profiles in the final experiment.

\subparagraph{State-of-the-art.} Blumenstock~\cite{doi:10.1137/1.9781611974317.10} provided us with a Java code implementing Kowalik's exact algorithm.
We ported that implementation with Dinic's algorithm to \cpp and validated that our implementation is on average 1.42 times faster than the Java implementation (compiled with OpenJDK~17 and run sequentially) on the instances used by Blumenstock~\cite{doi:10.1137/1.9781611974317.10}. 
A more detailed comparison of the running time of these implementations is reported
in Appendix Table~\ref{tab:javacomp}.
We have also implemented Georgakopoulos and Politopoulos's approach to be able to do a more conclusive comparison than what was available in~\cite{doi:10.1137/1.9781611974317.10}.
We can run this algorithm with Dinic's algorithm or Push-Relabel algorithm for better performance.
We use a Push-Relabel implementation in our final experiments for comparison since it is 1.6 times faster than Dinic's algorithm on our implementation.
The algorithm by Georgakopoulos and Politopoulos computes only  the pseudoarboricity, i.e., the objective value of an orientation, and not an orientation itself.
 As described by Blumenstock~\cite{doi:10.1137/1.9781611974317.10} our implementation computes the pseudoarboricity and uses Kowalik's reorientation scheme once to obtain an orientation.
In the following, we refer to our implementation of Georgakopoulos and Politopoulos~\cite{georgakopoulos2007max} by \COMPETITORGP and to our implementation of the exact Kowalik~\cite{kowalik2006approximation}~by~\COMPETITORKOWALIK.
Preliminary experiments on the test set showed that using the 2-approximation initialization before \COMPETITORKOWALIK and  \COMPETITORGP results in, respectively, $1.90$ and $3.02$ times faster running time than not using this initialization.
Therefore, we run the final experiment with 2-approximation for both of the state-of-the-art methods.
We note that our experiments with \COMPETITORGP and \COMPETITORKOWALIK are more comprehensive and conclusive than what were available in earlier work~\cite{doi:10.1137/1.9781611974317.10}.

\subparagraph{Implemented Algorithms.}
 For our main approaches we implemented a vanilla DFS (\ALGODFSNAIVE) and the algorithms  devised in Section~\ref{sec:engineering}.
 These include Fast Improvement (\ALGOFASTIMPROVE), DFS with improvements (\ALGODFS), BFS (\ALGOBATCHEDBFS) and the eager path search (\ALGOEAGERPATHSEARCH).
  Our implementation of  $2$-approximation (\ALGOTWOAPPROX) and the $2$-approximation conditioned on the average density (\ALGOTWOAPPROXCONDITIONAL)  can be run before all algorithms~considered~here.

\subsection{Parameter Study for the Proposed Algorithms}
In the subsequent sections, we adjust our parameters and investigate the different algorithmic components described in this paper using the nine graphs selected specifically for parameter tuning. 
\subparagraph{Naive Methods.} 
We conducted a comparative analysis of several DFS implementations: one without any of the techniques outlined in Section~\ref{sec:path:find:alg} (\ALGODFSNAIVEWITHOUTSHARED), one that incorporates all the proposed techniques (\ALGODFS), another that reuses the visited array (\ALGODFSNAIVESHARED), and the batched BFS (\ALGOBATCHEDBFS). All these implementations utilized the \ALGOFASTIMPROVE approach, and we present only the average execution times.
Our findings indicate that the \ALGODFS method is the most efficient, being significantly faster than the others. Specifically, the \ALGODFSNAIVESHARED is 2.02 times slower, and the \ALGOBATCHEDBFS is notably slower by a factor of 10.38. The slowest was the \ALGODFSNAIVEWITHOUTSHARED, which was 206.443 times slower than the optimized \ALGODFS. Based on these results, we will exclusively use the optimized \ALGODFS, i.e.~DFS with improvements, for the subsequent parameter tuning experiments.

\subparagraph{Fast Improve.} 
To demonstrate the effectiveness of the \ALGOFASTIMPROVE technique, we conducted experiments on the small dataset, running \ALGODFS both with and without the technique. The variant lacking \ALGOFASTIMPROVE failed to complete within 5 hours on the  instance from the MAWI set, which notably has an extremely high maximum degree of approximately \numprint{63}M.
                In stark contrast, the \ALGOFASTIMPROVE-enabled variant completed the same instance in just 18.84 seconds. Across the remaining eight instances, the non-\ALGOFASTIMPROVE variant performed consistently slower, averaging 27\% longer processing times.
Given that omitting \ALGOFASTIMPROVE led to a timeout on one instance and consistently slower performance on others, we use the \ALGOFASTIMPROVE algorithm for all instances.

 \subparagraph*{Eager Path Search.}
 We now optimize the parameters for our eager path search.
 In the eager path search we search for improving paths not only for the peak vertices, but for layers with lower out-degree.
 The parameters are the number of layers eagerly searched and the maximum size of a layer that will be searched for even more eagerly.
 Furthermore, we probe which method (\ALGOBATCHEDBFS~ or  \ALGODFS) should be used to finalize the orientation.

\begin{figure}[t]
  \centering
      \caption{\vspace*{-10pt}} 
      \begin{subfigure}[t]{0.49\textwidth}
        \centering
          \caption{ The running time of \ALGOEAGERPATHSEARCHPLUSDFS and \ALGOEAGERPATHSEARCHPLUSBFS normalized by \ALGOEAGERPATHSEARCHPLUSDFS plotted against the resulting min max out-degree. Low out-degree solutions are solved faster by \ALGODFS as finishing method.}
        \includegraphics{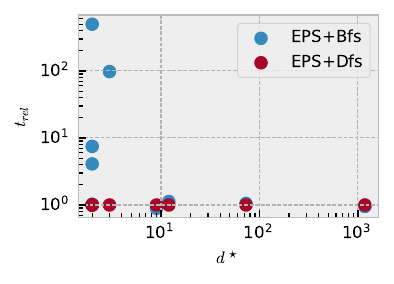}
          \label{fig:scatter:minout:runningtime}
      \end{subfigure}\hfill
      \begin{subfigure}[t]{0.49\textwidth}
        \centering
          \caption{ The  running time of \ALGOEAGERPATHSEARCH with or without \hbox{$2$-approximation} plotted against the average density $\rho=\frac{m}{n}$ of~the~graph (normalized by \ALGOEAGERPATHSEARCH).}
        \vspace{0.275cm}
        \includegraphics{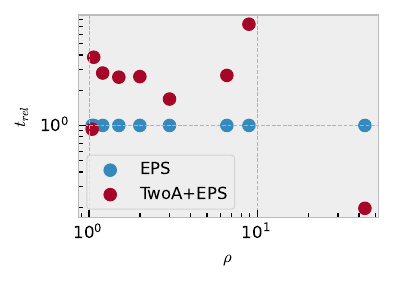}
          \label{fig:scatter:2approx:runningtime}
      \end{subfigure}
      \vspace*{-.5cm}
  \end{figure}

We tested the number of layers eagerly searched for values $i\in\{2,5,10, \sqrt{\max d -\rho}\}$ with $\rho=\dfrac{m}{n}$ and $\max d$ being the current maximum out-degree of the starting orientation. %
 We used the optimized \ALGODFS to compute the final orientation.
 For the sampled set the dynamic approach $i=\sqrt{\max d -\rho}$ yields the best results on average. The static approach with $i=5$ is only $2$ \% slower, while the approaches with $i=2$ and $10$ are $7.3$ resp. $8.3$ \% slower.
 We select the dynamic approach for our final~experiment and now progress into the second parameter the eager size.
If a layer size is smaller than the eager size the algorithm tries to find multiple improvements consecutively, instead of requiring the vertices to be collected again. 
In this experiment we combined the dynamic approach and DFS with the eager size parameter $c\in\{10,100,1000,10000\}$.
 The best geometric mean we can report for $c=100$,  for $c=10,1000,10000$ we report a  $1.5$\%, $0.6$\% resp. $0.16$\% higher geometric mean for the running time.
Therefore, we select $c=100$ for our final experiment. %
Finally, we investigate when to run either the \ALGOBATCHEDBFS~or~\ALGODFS~ after the eager path search.
Figure~\ref{fig:scatter:minout:runningtime} shows the running time divided by the optimal value plotted against the min max out-degree.
 We observe that for lower out-degree ($<10$) it is more beneficial to run the optimized \ALGODFS.
On one instance with low out-degree using \ALGOBATCHEDBFS is 60 times slower.
However, on instances with higher out-degree the~\ALGOBATCHEDBFS is consistently faster by a small relative margin.
We run  \ALGODFS, if the out-degree is less than 10, and~otherwise~\ALGOBATCHEDBFS.

\subparagraph{Two Approximation.} 

We now look into the efficiency of the $2$-approximation as outlined in Section~\ref{sec:2approx}. Specifically, we investigate the effectiveness of not always executing the two-approximation prior to the eager path search. The outcomes of this study are illustrated in Figure~\ref{fig:scatter:2approx:runningtime}, where we plot the average density of the graphs against their average running time. The experiment indicates that for graphs with low average density, pre-running the two-approximation does not yield beneficial results.
  We run the $2$-approximation exclusively for graphs with an average density exceeding 10 to achieve reasonable reductions on such graphs. %

\subparagraph{Final Algorithm.} The final configuration of our algorithm \ALGOFINALNAME (RPO) is as follows: After using the $2$-approximation conditionally on the average density ($\frac{m}{n}>10$) and executing the \ALGOFASTIMPROVE algorithm, it runs the \ALGOEAGERPATHSEARCH with dynamic layer count of $\sqrt{(\max d -\rho)}$. If less than $100$ vertices are in a layer they are relaxed more eagerly.
Finally, in order to produce a correct result a \ALGOBATCHEDBFS~is run for out-degree higher than $10$ and the specialized \ALGODFS for~lower~degree.

\vspace{-0.25cm}
\subsection{Comparison with State-of-the-Art}
\begin{figure}[t]
  \centering
  \vspace*{-.5cm}
  \caption{Running time performance profile on 58 instances for state-of-the-art algorithms \COMPETITORKOWALIKTwoApprox,~\COMPETITORGPTwoApprox~ and our \ALGOFINALNAME~(\textit{RPO}) approach. On the right hand side we present a detailed zoom to values up to 6.5.}%
  \vspace{-0.25cm}
  \includegraphics{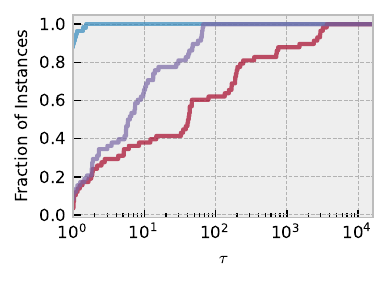}
  \includegraphics{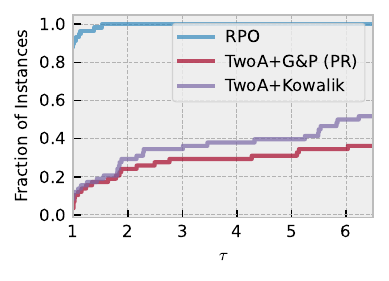}
  \label{fig:time:perf:profile}
  \vspace*{-0.5cm}
\end{figure}
We now compare our approach to the state-of-the-art on the whole dataset.
In Figure~\ref{fig:time:perf:profile} the performance profile for   our algorithm (\ALGOFINALNAME)  is shown in comparison to the algorithm by Kowalik~\cite{kowalik2006approximation}  (\COMPETITORKOWALIKTwoApprox) and Georgakopoulos \& Politopoulos~\cite{georgakopoulos2007max} (\COMPETITORGPTwoApprox) with combined with the $2$--approximation on  58 problem instances. %
On average our algorithm is 6.59 times faster than \COMPETITORKOWALIKTwoApprox and 36.27 times faster than \COMPETITORGPTwoApprox; moreover on 88 \% of instances it is the fastest approach.
As seen in the right plot of Figure~\ref{fig:time:perf:profile}, \ALGOFINALNAME solves all instances within a factor of $\tau<1.56$ of the fastest algorithm (on an instance) and the competitor approaches solve around 50 \% of instances within a factor of $\tau=6$.
 Since the profile of \ALGOFINALNAME is higher than both  \COMPETITORKOWALIKTwoApprox and \COMPETITORGPTwoApprox for all values of $\tau$, we conclude that
the proposed approach is faster than the reimplementation of the state-of-the-art methods.
Moreover, as shown in the preliminary experiment, the speed up over the Java implementations is even more pronounced.
We report detailed per instance results of the running times in Appendix Table~\ref{table:results}. Table~\ref{table:groupedresults} shows the running times averaged per sub dataset. We now give a more detailed discussion of the performance of the different algorithm for the subgroups of our dataset. %

In general, it can be observed that competing algorithms can solve instances from the MAWI subset faster. 
Out of all the instances for which our approach is not the fastest, our algorithm experiences the highest relative slowdown on the \HRinstance{mawi\_'130} instance. Here, \ALGOFINALNAME needs 56\% more time than both competing algorithms. %
This is due to the fact that the $2$-approximation is  very effective on the MAWI dataset, however the $2$=approximation is not run by our algorithm on these instances as they have a low average density ($\frac{m}{n}<2$).
 When running our approach on these instances with the $2$-approximation algorithm as a preprocessing algorithm, \hbox{we achieve similar running times.}

The performance differences are very big especially on the low density and low out-degree instances of the \HRdataset{flowipm22} set. 
For example, on the \HRinstance{spielman\_k600} instance  \COMPETITORGPTwoApprox is 243.10 times and  \COMPETITORKOWALIKTwoApprox is 65.46 times slower than our new approach.
In general, this is due to the fact that the FlowIPM22 instances have a high maximum degree compared to their optimum out-degree of $3$.
 The competitors can not limit their search space properly and have to perform  many reorientations using their flow~algorithm.

 Similarly, we can observe advantages of our approach on random generated graphs  in the \HRdataset{dimacs10} set, like the \HRinstance{delaunay\_n24} instance. \ALGOFINALNAME is around 80 times faster than  \COMPETITORKOWALIKTwoApprox and 221 times faster than \COMPETITORGPTwoApprox.
Moreover, the \COMPETITORGPTwoApprox algorithm has issues with some instances from numerical backgrounds (\HRinstance{huge*}) in the dimacs10 instances, resulting in a running time much higher than the other approaches (up to 16393.74 times slower).
 The \COMPETITORGPTwoApprox algorithm has to solve more flows  to converge according to its binary search scheme, while its pruning mechanism is not that efficient on these instances.
 The pruning removes vertices with unsaturated edges connected to the target vertex, when a new lower bound for the pseudoarboricity is accepted, which does not happen often on those instances.
 In general,  our new algorithm \ALGOFINALNAME solves instances from numerical backgrounds and road networks, represented in the dimacs10 set, significantly faster than the previous~approaches.
\begin{wrapfigure}{r}{0.5\linewidth}
  \begin{minipage}{\linewidth}
    \vspace{-0.75cm}
  \begin{table}[H]
  \caption{Geometric mean running time in seconds  grouped by sub data set and for all instances.
  }
  \centering
  \label{table:groupedresults}
\begin{tblr}[
caption = {Geometric mean running time in seconds  grouped by sub data set and for all instances.
},
  label = {},
 ]{
  colsep = 1mm,
  rowsep = 1mm,
  colspec = {lS[table-format=5.2]S[table-format=4.2]S[table-format=4.2]},rowhead =1, hlines,  row{even} = {gray9}, row{1} = {lipicsYellow,font=\bfseries},cells   = {font = \footnotesize},}
  &{{{\COMPETITORGPTwoApprox}}}&{{{\COMPETITORKOWALIKTwoApprox}}}&{{{RPO}}}\\
  flowipm22~\cite{doi:10.1137/19M123806X} &104.62&33.57&\bfseries 0.57\\
  gap~\cite{DBLP:journals/corr/BeamerAP15} &12267.57&4808.19&\bfseries 2628.25\\
  genbank~\cite{benson1996genbank} &70.14&70.63&\bfseries 36.69\\
  law~\cite{BoVWFI,BRSLLP} &5.02&2.97&\bfseries 1.82\\
  mawi~\cite{cho2000traffic} &8.40&\bfseries 8.39&10.07\\
  snap~\cite{snap} &8.86&5.15&\bfseries 1.85\\
  sybrandt~\cite{DBLP:conf/cikm/SybrandtTSS20} &5340.81&3064.19&\bfseries 2226.32\\
  dimacs10~\cite{benchmarksfornetworksanalysis} &337.96&21.49&\bfseries 1.86\\
\hline
\bfseries all&144.85 &26.31 &\bfseries 3.99 
  \end{tblr}
\end{table}

  \vspace{-0.5cm}
\end{minipage} 
\end{wrapfigure}
  \indent On the first instance  from the \HRdataset{gap} dataset, \HRinstance{gap-urand},   \COMPETITORGPTwoApprox requires 81481.60 seconds to finish, 15.71-fold of what our new approach and 6.31-fold of what \COMPETITORKOWALIKTwoApprox require to finish. On the \HRinstance{gap-kron} instance, both competitors are 1.38 (\COMPETITORGPTwoApprox) resp. 1.34 (\COMPETITORKOWALIKTwoApprox) times slower.
  On all five instances from \HRdataset{genbank} set, our new approach is about 1.9 times faster than each of the competitors. There are no significant deviations in this dataset.
   On the single instance from the \HRdataset{law} dataset we  report that \ALGOFINALNAME is 1.63 times faster than \COMPETITORKOWALIKTwoApprox and 2.76 times faster than \COMPETITORGPTwoApprox.
   The \HRdataset{snap} dataset contains social graphs and road networks. On the three road networks our approach requires less than 1/10th of the running time in comparison to the fastest competitor (\COMPETITORKOWALIKTwoApprox).
    For two  of the social graphs (\HRinstance{com-youtube},\HRinstance{as-skitter})  all approaches require nearly exactly the same time.
    The \HRinstance{com-friendster}  instance is solved by \COMPETITORGPTwoApprox about 1.13 times faster than our approach.
   On the  \HRinstance{com-livejournal} the \COMPETITORKOWALIKTwoApprox is faster by a factor of 1.08.

   The sybrandt set contains the two biggest instances in our dataset.
   For the \HRinstance{moliere'16} \COMPETITORKOWALIKTwoApprox is the fastest algorithm, our approach is 14\% slower.
  \ALGOFINALNAME solves the \HRinstance{agatha'15} instance 2.16 times faster than the fastest competitor \COMPETITORKOWALIKTwoApprox.

\section{Conclusion}\label{sec:conclusion}
We have proposed a new framework for algorithms solving the edge orientation problem based on the ideas of Venkateswaran~\cite{venkateswaran2004minimizing} and gave a new flow-based proof.
We have investigated a vast variety of engineering techniques for the problem.
 Our techniques include a fast improvement heuristic, specialized depth-first-search, scheduling path searches more eagerly as well as running a data reduction based on a 2-approximation algorithm. 
Experiments have shown that with our final algorithm outperforms the fastest exact state-of-art algorithm by a factor of 6.59 on average.
 Especially on low density and low out-degree instances, like road networks and instances from numerical backgrounds, our algorithm outperforms its competitors.
 Only on low density, high out-degree instances the competitors have an advantage and compute an orientation slightly faster.

There are multiple areas of future work. 
Most importantly, we think that an explicit parallelism for the proposed algorithms is worthwhile as on massive instances we still observe very large running times overall.
\vfill \pagebreak

\bibliography{all_clean}
\begin{appendix}
\ifCompileTables
\section{Additional Tables}
  \begin{longtblr}[
  caption = {Running times [s] comparing \textsf{C}\texttt{++} with Java implementation of exact Kowalik~\cite{kowalik2006approximation} by Blumenstock~\cite{doi:10.1137/1.9781611974317.10} on the data set presented in \cite{doi:10.1137/1.9781611974317.10}.
  },
    label = {tab:javacomp},
   ]{
  colspec = {llSS},rowhead =1, hlines,  row{even} = {gray9}, row{1} = {lipicsYellow,font=\bfseries},cells   = {font = \footnotesize}, column{1}={font=\ttfamily}}
  &$d^\star$&{{{Java}}}&{{{\textsf{C}\texttt{++}}}}\\
  com-Amazon&\numprint{5}&2.47&\bfseries 1.81\\
  com-Youtube&\numprint{46}&14.96&\bfseries 8.15\\
  com-DBLP&\numprint{57}&1.70&\bfseries 1.15\\
  com-LiveJournal&\numprint{194}&86.85&\bfseries 70.19\\
  com-Orkut&\numprint{228}&382.50&\bfseries 303.49\\
  \end{longtblr}

\begin{longtblr}[
  caption = {Average statistics for the sub data sets. \# = number of instances in data set $+$ number of instances in test set.},
    label = {table:stats:per:group},
   ]{
  colspec = {lrS[scientific-notation = true,round-mode = figures,
  round-precision = 2,
  table-format=1.2e1
]S[scientific-notation = true,round-mode = figures,
round-precision = 2,
table-format=1.2e1
]S[scientific-notation = true,round-mode = figures,
round-precision = 2,
table-format=1.0e1
]S[scientific-notation = true,round-mode = figures,
round-precision = 2,
table-format=1.2e1
]S},rowhead =1, hlines,font={\footnotesize}, row{even} = {gray9}, row{1} = {lipicsYellow,font=\bfseries},cells   = {font = \footnotesize}}
 &\#&{{{$n$}}}&{{{$m$}}}&{{{components}}}&{{{$\max d$}}}&{{{$\min d$}}}\\
 flowipm22~\cite{doi:10.1137/19M123806X}&5&29423602.00&29514001.00& 1.00&800.00&2.00\\
 gap~\cite{DBLP:journals/corr/BeamerAP15}&2&134217727.00&2129557846.00&35582132.00&786453.00& 3.00\\
 genbank~\cite{benson1996genbank} &5&129369000.60&137982222.20&14556.20&32.40& 1.00\\
 law~\cite{BoVWFI,BRSLLP}&1&1139905.00&56375711.00&44508.00&11467.00& 0.00\\
 mawi~\cite{cho2000traffic}&4&102331852.75&107851058.75&1829168.25&94717138.00& 0.75\\
 snap~\cite{snap}&8&9995707.25&247280811.75&3299.62&14698.00& 0.62\\
 sybrandt~\cite{DBLP:conf/cikm/SybrandtTSS20}&2&107101882.00&4566506832.50&12519.50&7374767.50& 0.50\\
 dimacs10~\cite{benchmarksfornetworksanalysis}&31&9428837.84&23094605.94&8175.45&4256.71& 1.61\\
 test set&9&13379543.11&30884952.22&204228.00&7028258.67& 1.78\\
  \end{longtblr}
\begin{adjustwidth}{-0cm}{-0cm}
  {
  { \begin{longtblr}[
caption = {Instance statistics for the 67 graphs used in experiments.$\sigma$ Relative size of 2-approximation, $\Delta$~maximum degree,$\delta$~minimum degree. Instances in the test set are \textbf{bold}.},
  label = {tab:instancedetails},
 ]{
  colspec = {llS[scientific-notation = true,round-mode = figures,
  round-precision = 2,
  table-format=1e2
  ]rS[scientific-notation = true,round-mode = figures,
  round-precision = 2,
  table-format=1e1
  ]rrS[scientific-notation = true,round-mode = figures,
  round-precision = 2,
  table-format=1.0e1
  ]S[scientific-notation = true,round-mode = figures,
  round-precision = 2,
  table-format=1e1
  ]r},rowhead =1, hlines,  row{even} = {gray9}, row{1} = {lipicsYellow,font=\bfseries}, column{1}={font=\ttfamily},cells   = {font = \footnotesize},column{1}={font=\ttfamily}}
  &&{{{n}}}&$\sigma_n$&{{{m}}}&$\sigma_m$&m/n&{{{comp.}}}&{{{$\Delta$}}}&$\delta$\\
  333sp&dimacs10 &\justprint{3712815}& 1.00&\justprint{11108633}&1.00&2.99&\justprint{1}&\justprint{28}&\justprint{2}\\
  \bfseries adaptive&dimacs10&\justprint{6815744}& \justprint{1}&\justprint{13624320}& \justprint{1}&2.00& \justprint{1}&\justprint{4}&\justprint{2}\\
  agatha'15&sybrandt &\justprint{183964077}& 0.03&\justprint{5794362982}&0.10&31.50&\justprint{13}&\justprint{12642631}&\justprint{1}\\
  as365&dimacs10 &\justprint{3799275}& 1.00&\justprint{11368076}&1.00&2.99&\justprint{1}&\justprint{14}&\justprint{2}\\
  \bfseries asia\_osm&dimacs10&\justprint{11950757}& 0.75&\justprint{12711603}&0.76&1.06&\justprint{1}&\justprint{9}&\justprint{1}\\
  \bfseries  channel-'b050&dimacs10&\justprint{4802000}& 1.00&\justprint{42681372}&1.00&8.89&\justprint{1}&\justprint{18}&\justprint{6}\\
  as-skitter&snap &\justprint{1696415}& 0.00&\justprint{11095298}&0.01&6.54&\justprint{756}&\justprint{35455}&\justprint{1}\\
  belgium\_osm&dimacs10 &\justprint{1441295}& 0.86&\justprint{1549970}&0.87&1.08&\justprint{1}&\justprint{10}&\justprint{1}\\
  com-friendster&snap &\justprint{65608366}& 0.00&\justprint{1806067135}&0.01&27.53&\justprint{1}&\justprint{5214}&\justprint{1}\\
  com-livejournal&snap &\justprint{3997962}& 0.00&\justprint{34681189}&0.01&8.67&\justprint{1}&\justprint{14815}&\justprint{1}\\
  com-orkut&snap &\justprint{3072441}& 0.01&\justprint{117185083}&0.05&38.14&\justprint{1}&\justprint{33313}&\justprint{1}\\
  com-youtube&snap &\justprint{1134890}& 0.00&\justprint{2987624}&0.03&2.63&\justprint{1}&\justprint{28754}&\justprint{1}\\
  delaunay\_n20&dimacs10 &\justprint{1048576}& \justprint{1}&\justprint{3145686}& \justprint{1}&3.00& \justprint{1}&\justprint{23}&\justprint{3}\\
  delaunay\_n21&dimacs10 &\justprint{2097152}& \justprint{1}&\justprint{6291408}& \justprint{1}&3.00& \justprint{1}&\justprint{23}&\justprint{3}\\
  delaunay\_n22&dimacs10 &\justprint{4194304}& \justprint{1}&\justprint{12582869}& \justprint{1}&3.00& \justprint{1}&\justprint{23}&\justprint{3}\\
  delaunay\_n23&dimacs10 &\justprint{8388608}& \justprint{1}&\justprint{25165784}& \justprint{1}&3.00& \justprint{1}&\justprint{28}&\justprint{3}\\
  delaunay\_n24&dimacs10 &\justprint{16777216}& \justprint{1}&\justprint{50331601}& \justprint{1}&3.00& \justprint{1}&\justprint{26}&\justprint{3}\\
  europe\_osm&dimacs10 &\justprint{50912018}& 0.80&\justprint{54054660}&0.81&1.06&\justprint{1}&\justprint{13}&\justprint{1}\\
  gap-kron&gap &\justprint{134217726}&0.00&\justprint{2111632322}&0.10&15.73&\justprint{71164263}&\justprint{1572838}& \justprint{0}\\
  gap-urand&gap &\justprint{134217728}& 1.00&\justprint{2147483370}&1.00&16.00&\justprint{1}&\justprint{68}&\justprint{6}\\
  germany\_osm&dimacs10 &\justprint{11548845}& 0.81&\justprint{12369181}&0.82&1.07&\justprint{1}&\justprint{13}&\justprint{1}\\
  g'b'\_osm&dimacs10 &\justprint{7733822}& 0.70&\justprint{8156517}&0.72&1.05&\justprint{1}&\justprint{8}&\justprint{1}\\
  hollywood-2009&law &\justprint{1139905}&0.00&\justprint{56375711}&0.06&49.46&\justprint{44508}&\justprint{11467}& \justprint{0}\\
  hugebubbles'00&dimacs10 &\justprint{18318143}& \justprint{1}&\justprint{27470081}& \justprint{1}&1.50& \justprint{1}&\justprint{3}&\justprint{2}\\
  hugebubbles'10&dimacs10 &\justprint{19458087}& \justprint{1}&\justprint{29179764}& \justprint{1}&1.50& \justprint{1}&\justprint{3}&\justprint{2}\\
  hugebubbles'20&dimacs10 &\justprint{21198119}& \justprint{1}&\justprint{31790179}& \justprint{1}&1.50& \justprint{1}&\justprint{3}&\justprint{2}\\
  hugetrace'00&dimacs10 &\justprint{4588484}& \justprint{1}&\justprint{6879133}& \justprint{1}&1.50& \justprint{1}&\justprint{3}&\justprint{2}\\
  hugetrace'10&dimacs10 &\justprint{12057441}& \justprint{1}&\justprint{18082179}& \justprint{1}&1.50& \justprint{1}&\justprint{3}&\justprint{2}\\
  hugetrace'20&dimacs10 &\justprint{16002413}& \justprint{1}&\justprint{23998813}& \justprint{1}&1.50& \justprint{1}&\justprint{3}&\justprint{2}\\
  hugetric'00&dimacs10 &\justprint{5824554}& \justprint{1}&\justprint{8733523}& \justprint{1}&1.50& \justprint{1}&\justprint{3}&\justprint{2}\\
  \bfseries hugetric'10&dimacs10&\justprint{6592765}& \justprint{1}&\justprint{9885854}& \justprint{1}&1.50& \justprint{1}&\justprint{3}&\justprint{2}\\
  hugetric'20&dimacs10 &\justprint{7122792}& \justprint{1}&\justprint{10680777}& \justprint{1}&1.50& \justprint{1}&\justprint{3}&\justprint{2}\\
  italy\_osm&dimacs10 &\justprint{6686493}& 0.80&\justprint{7013978}&0.81&1.05&\justprint{1}&\justprint{9}&\justprint{1}\\
  kmer\_a2a&genbank &\justprint{170728175}& 0.00&\justprint{180292586}&0.00&1.06&\justprint{5353}&\justprint{40}&\justprint{1}\\
  kmer\_p1a&genbank &\justprint{139353211}& 0.00&\justprint{148914992}&0.00&1.07&\justprint{7944}&\justprint{40}&\justprint{1}\\
  kmer\_u1a&genbank &\justprint{67716231}& 0.00&\justprint{69389281}&0.00&1.02&\justprint{44132}&\justprint{35}&\justprint{1}\\
  kmer\_v1r&genbank &\justprint{214005017}& 0.00&\justprint{232705452}&0.00&1.09&\justprint{9}&\justprint{8}&\justprint{1}\\
  kmer\_v2a&genbank &\justprint{55042369}& 0.00&\justprint{58608800}&0.00&1.06&\justprint{15343}&\justprint{39}&\justprint{1}\\
  kron\_g500-'20&dimacs10 &\justprint{1048576}&0.01&\justprint{44619402}&0.13&42.55&\justprint{253380}&\justprint{131503}& \justprint{0}\\
  \bfseries kron\_g500-'21&dimacs10&\justprint{2097152}&0.00&\justprint{91040932}&0.10&43.41&\justprint{553159}&\justprint{213904}& \justprint{0}\\  
  mawi'000&mawi &\justprint{35991342}& 0.00&\justprint{37242710}&0.00&1.03&\justprint{746178}&\justprint{32481338}&\justprint{1}\\
  \bfseries mawi'030&mawi &\justprint{68863315}& 0.00&\justprint{71707480}&0.00&1.04&\justprint{1284884}&\justprint{63040326}&\justprint{1}\\
  mawi'130&mawi &\justprint{128568730}& 0.00&\justprint{135117420}&0.00&1.05&\justprint{2156828}&\justprint{119191841}&\justprint{1}\\
  mawi'330&mawi &\justprint{226196185}&0.00&\justprint{240023945}&0.00&1.06&\justprint{3971144}&\justprint{210795477}& \justprint{0}\\
  mawi'345&mawi &\justprint{18571154}& 0.00&\justprint{19020160}&0.00&1.02&\justprint{442523}&\justprint{16399896}&\justprint{1}\\
  moliere'16&sybrandt &\justprint{30239687}&0.00&\justprint{3338650683}&0.00&110.41&\justprint{25026}&\justprint{2106904}& \justprint{0}\\
  m6&dimacs10 &\justprint{3501776}& \justprint{1}&\justprint{10501936}& \justprint{1}&3.00& \justprint{1}&\justprint{10}&\justprint{3}\\
  naca0015&dimacs10 &\justprint{1039183}& \justprint{1}&\justprint{3114818}& \justprint{1}&3.00& \justprint{1}&\justprint{10}&\justprint{3}\\
  n'l'\_osm&dimacs10 &\justprint{2216688}& 0.83&\justprint{2441238}&0.84&1.10&\justprint{1}&\justprint{7}&\justprint{1}\\
  \bfseries nlr&dimacs10&\justprint{4163763}& \justprint{1}&\justprint{12487976}& \justprint{1}&3.00& \justprint{1}&\justprint{20}&\justprint{3}\\
  packing-'b050&dimacs10 &\justprint{2145852}&0.94&\justprint{17488243}&0.95&8.15&\justprint{14}&\justprint{18}& \justprint{0}\\
  \bfseries rgg\_n\_2\_20\_s0&dimacs10&\justprint{1048576}&0.02&\justprint{6891620}&0.02&6.57&\justprint{3}&\justprint{36}& \justprint{0}\\
  rgg\_n\_2\_21\_s0&dimacs10 &\justprint{2097152}&0.27&\justprint{14487995}&0.30&6.91&\justprint{8}&\justprint{37}& \justprint{0}\\
  rgg\_n\_2\_22\_s0&dimacs10 &\justprint{4194304}&0.14&\justprint{30359198}&0.15&7.24&\justprint{5}&\justprint{36}& \justprint{0}\\
  rgg\_n\_2\_23\_s0&dimacs10 &\justprint{8388608}&0.07&\justprint{63501393}&0.07&7.57&\justprint{5}&\justprint{40}& \justprint{0}\\
  rgg\_n\_2\_24\_s0&dimacs10 &\justprint{16777216}&0.03&\justprint{132557200}&0.04&7.90&\justprint{2}&\justprint{40}& \justprint{0}\\
  \bfseries road\_central&dimacs10&\justprint{14081816}& 0.71&\justprint{16933413}&0.76&1.20&\justprint{1}&\justprint{8}&\justprint{1}\\
  roadnet-ca&snap &\justprint{1971281}&0.81&\justprint{2766607}&0.87&1.40&\justprint{8713}&\justprint{12}& \justprint{0}\\
  roadnet-pa&snap &\justprint{1090920}&0.80&\justprint{1541898}&0.86&1.41&\justprint{3034}&\justprint{9}& \justprint{0}\\
  roadnet-tx&snap &\justprint{1393383}&0.78&\justprint{1921660}&0.85&1.38&\justprint{13890}&\justprint{12}& \justprint{0}\\
  road\_usa&dimacs10 &\justprint{23947347}& 0.71&\justprint{28854312}&0.76&1.20&\justprint{1}&\justprint{9}&\justprint{1}\\
  spielman\_k200&flowipm22 &\justprint{2686802}& \justprint{1}&\justprint{2707001}& \justprint{1}&1.01& \justprint{1}&\justprint{400}&\justprint{2}\\
  spielman\_k300&flowipm22 &\justprint{9045202}& \justprint{1}&\justprint{9090501}& \justprint{1}&1.01& \justprint{1}&\justprint{600}&\justprint{2}\\
  spielman\_k400&flowipm22 &\justprint{21413602}& \justprint{1}&\justprint{21494001}& \justprint{1}&1.00& \justprint{1}&\justprint{800}&\justprint{2}\\
  spielman\_k500&flowipm22 &\justprint{41792002}& \justprint{1}&\justprint{41917501}& \justprint{1}&1.00& \justprint{1}&\justprint{1000}&\justprint{2}\\
  spielman\_k600&flowipm22 &\justprint{72180402}& \justprint{1}&\justprint{72361001}& \justprint{1}&1.00& \justprint{1}&\justprint{1200}&\justprint{2}\\
  venturilevel3&dimacs10 &\justprint{4026819}& 0.83&\justprint{8054237}&0.83&2.00&\justprint{1}&\justprint{6}&\justprint{2}\\
  \end{longtblr}}
 
\begin{longtblr}[
caption = {Average running time per instance. Sorted by $d^\star$. $5$ repetitions.},
  label = {table:results},
 ]{
colspec = {lrS[table-format=10.2]S[table-format=10.2]S[table-format=10.2]},rowhead =1, hlines,  row{even} = {gray9}, row{1} = {lipicsYellow,font=\bfseries}, column{1}={font=\ttfamily}}
&$d^\star$&{{{\COMPETITORGPTwoApprox}}}&{{{\COMPETITORKOWALIKTwoApprox}}}&{{{\ALGOFINALNAME}}}\\
adaptive&\numprint{2}&5508.26&8.80&\bfseries 1.40\\
asia\_osm&\numprint{2}&56.45&14.28&\bfseries 1.16\\
belgium\_osm&\numprint{2}&7.17&1.72&\bfseries 0.17\\
europe\_osm&\numprint{2}&288.53&88.28&\bfseries 6.65\\
germany\_osm&\numprint{2}&61.71&13.60&\bfseries 1.62\\
g'b'\_osm&\numprint{2}&33.52&7.72&\bfseries 0.96\\
hugebubbles'00&\numprint{2}&70001.31&31.00&\bfseries 4.27\\
hugebubbles'10&\numprint{2}&31454.70&44.19&\bfseries 8.66\\
hugebubbles'20&\numprint{2}&27227.08&46.97&\bfseries 8.49\\
hugetrace'00&\numprint{2}&2901.07&7.22&\bfseries 0.98\\
hugetrace'10&\numprint{2}&8391.65&22.87&\bfseries 3.50\\
hugetrace'20&\numprint{2}&7916.97&31.78&\bfseries 5.27\\
hugetric'00&\numprint{2}&3568.83&9.04&\bfseries 1.22\\
hugetric'10&\numprint{2}&4805.79&13.60&\bfseries 2.46\\
hugetric'20&\numprint{2}&7161.24&14.68&\bfseries 2.67\\
italy\_osm&\numprint{2}&31.45&8.07&\bfseries 0.76\\
n'l'\_osm&\numprint{2}&10.50&2.15&\bfseries 0.22\\
roadnet-ca&\numprint{2}&11.33&3.44&\bfseries 0.25\\
roadnet-pa&\numprint{2}&5.71&1.72&\bfseries 0.14\\
road\_central&\numprint{2}&92.38&26.98&\bfseries 4.60\\
road\_usa&\numprint{2}&145.56&34.94&\bfseries 3.56\\
spielman\_k200&\numprint{2}&14.16&5.42&\bfseries 0.08\\
spielman\_k300&\numprint{2}&51.19&16.52&\bfseries 0.38\\
spielman\_k400&\numprint{2}&123.57&39.95&\bfseries 0.64\\
spielman\_k500&\numprint{2}&250.26&79.07&\bfseries 1.24\\
spielman\_k600&\numprint{2}&559.14&150.55&\bfseries 2.30\\
333sp&\numprint{3}&2408.29&95.47&\bfseries 3.19\\
as365&\numprint{3}&2820.13&148.65&\bfseries 3.90\\
delaunay\_n20&\numprint{3}&72.27&19.73&\bfseries 0.47\\
delaunay\_n21&\numprint{3}&168.78&48.29&\bfseries 1.01\\
delaunay\_n22&\numprint{3}&413.96&123.88&\bfseries 2.20\\
delaunay\_n23&\numprint{3}&950.79&322.33&\bfseries 4.82\\
delaunay\_n24&\numprint{3}&2275.57&816.72&\bfseries 10.28\\
m6&\numprint{3}&1415.00&132.64&\bfseries 4.08\\
naca0015&\numprint{3}&613.64&24.18&\bfseries 0.86\\
kmer\_v1r&\numprint{3}&132.97&133.62&\bfseries 58.33\\
nlr&\numprint{3}&1500.28&176.09&\bfseries 4.70\\
roadnet-tx&\numprint{3}&7.13&2.26&\bfseries 0.15\\
venturilevel3&\numprint{3}&25.09&3.44&\bfseries 0.59\\
kmer\_u1a&\numprint{5}&40.80&40.71&\bfseries 21.67\\
kmer\_v2a&\numprint{8}&32.63&33.10&\bfseries 18.30\\
channel-'b050&\numprint{9}&1031.67&10.47&\bfseries 0.66\\
packing-'b050&\numprint{9}&76.12&4.54&\bfseries 0.41\\
kmer\_a2a&\numprint{10}&110.12&109.49&\bfseries 59.56\\
kmer\_p1a&\numprint{10}&87.13&89.17&\bfseries 48.26\\
rgg\_n\_2\_20\_s0&\numprint{12}&0.88&0.77&\bfseries 0.25\\
rgg\_n\_2\_21\_s0&\numprint{12}&7.86&3.46&\bfseries 0.59\\
rgg\_n\_2\_22\_s0&\numprint{13}&9.89&6.09&\bfseries 1.11\\
rgg\_n\_2\_23\_s0&\numprint{14}&13.85&9.49&\bfseries 2.22\\
rgg\_n\_2\_24\_s0&\numprint{14}&22.65&18.12&\bfseries 5.24\\
gap-urand&\numprint{17}&81481.60&12910.44&\bfseries 5183.51\\
mawi'345&\numprint{40}&2.26&2.29&\bfseries 2.08\\
com-youtube&\numprint{46}&0.71&0.58&\bfseries 0.58\\
mawi'000&\numprint{58}&4.41&\bfseries 4.38&4.48\\
mawi'030&\numprint{73}&\bfseries 8.16&8.20&9.67\\
mawi'130&\numprint{78}&\bfseries 14.67&14.67&22.85\\
as-skitter&\numprint{90}&1.08&\bfseries 1.03&1.04\\
mawi'330&\numprint{93}&33.95&\bfseries 33.65&48.18\\
agatha'15&\numprint{97}&22506.38&9267.07&\bfseries 4291.18\\
com-livejournal&\numprint{194}&4.26&\bfseries 4.14&4.51\\
com-orkut&\numprint{228}&27.61&14.00&\bfseries 9.18\\
moliere'16&\numprint{232}&1267.39&\bfseries 1013.19&1155.05\\
com-friendster&\numprint{274}&\bfseries 917.42&1056.45&1039.20\\
kron\_g500-'20&\numprint{908}&13.62&5.73&\bfseries 2.65\\
hollywood-2009&\numprint{1104}&5.02&2.97&\bfseries 1.82\\
kron\_g500-'21&\numprint{1178}&21.94&10.85&\bfseries 5.78\\
gap-kron&\numprint{2369}&1846.96&1790.70&\bfseries 1332.63\\
\end{longtblr}
}
\end{adjustwidth}
  \else
  \fi
\end{appendix}
\end{document}